\documentclass[runningheads]{svmult}
\usepackage{makeidx}   
\usepackage{graphicx}  
\usepackage{subeqnar}  
\usepackage{multicol}  
\usepackage{cropmark} 
\usepackage{physprbb}  
\makeindex             
\begin{document}
\renewcommand{\thefootnote}{\fnsymbol{footnote}}
\thispagestyle{empty}
\begin{flushright}
{\bf DESY 99-154}\\
{\bf hep-ph/9910329}\\
{\bf September 1999}\\[3cm]
\end{flushright}
\begin{center}
\begin{Large}
{\bf
Theory of Elastic Vector Meson Production}\footnote{Presented 
at the {\em Ringberg Workshop: New Trends in HERA Physics 1999}, Ringberg
Castle, Tegernsee, Germany, 30.~May -- 4.~June 1999.}\\[1.5cm]
\end{Large}
\begin{large}
Thomas Teubner\\[1.cm]
\end{large}
{\it
Deutsches Elektronen-Synchrotron DESY, Notkestrasse 85, D-22607 Hamburg,
Germany}\\
\end{center}
\renewcommand{\thefootnote}{\arabic{footnote}}
\vfill
\newpage

\mbox{}
\thispagestyle{empty}
\pagebreak

\setcounter{page}{1}
\title*{Theory of Elastic Vector Meson Production}
\author{Thomas Teubner
}
\institute{Deutsches Elektronen-Synchrotron DESY, Notkestrasse 85,
  D-22607 Hamburg, Germany}
\maketitle              

\begin{abstract}
The elastic production of vector mesons at HERA is discussed from the
theoretical point of view. We briefly review different models, their
successes and shortcomings. Main emphasis is put on recent issues in
perturbative QCD calculations. Models including the vector meson wave
function are compared with an approach based on parton-hadron
duality. We discuss several refinements of these models in some
detail, including the important role of off-diagonal parton distributions.
\end{abstract}

\section{Introduction}
Why are we interested in elastic vector meson production? First of all
the process $\gamma^* p \to V p$ provides us with well distinguishable
experimental signals in a wide range of the $\gamma^* p$ c.m. energy $W$, the
virtuality of the photon $Q^2$, and the mass of the vector meson
$M_V$. Quite some data are already available for $V = \rho, \phi$ and
$J/\Psi$, and even for the heavy $\Upsilon$ first measurements were
published recently.\footnote{For the discussion of experimental
  results on the production of light and heavy quarkonia see
  \cite{AProskuryakov,CKiesling,LLindemann} and references therein.}
In the future the range in $Q^2$ and $W$ and the precision of the data
will increase. This enables us to study vector meson production in
detail in the very interesting regime where the transition from soft
to hard QCD dynamics is expected (and already seen) to take place. In
addition, there is hope to make use of the high sensitivity of this
process on the gluon distribution $x\,g(x,\overline{Q}^2)$ in the proton to
constrain this quantity at small values of $x$ better than through
other processes.

In the following we first sketch the basic picture of elastic vector meson
production. In Section 2 we briefly discuss different theoretical
models which are not based on the two gluon exchange picture which is
then introduced in Section~3. There, starting from the basic leading
order result known for long time, we develop corrections which improve
the leading order formula. In Section 4 recent issues in pQCD
calculations as off-diagonal parton distributions, the influence of
the vector meson wave function and an alternative approach using
parton-hadron duality are discussed. We mainly concentrate on 
diffractive $\rho$ meson electroproduction, but the presented
perturbative model is also successful in the case of $J/\Psi$ and $\Upsilon$.
Section 4 contains our conclusions and outlook.

\subsection{The basic picture}
\begin{figure}[htb]
\begin{center}
\includegraphics[width=0.7\textwidth]{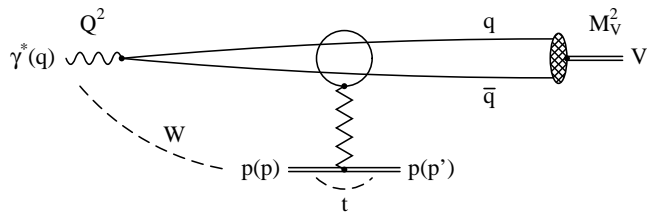}
\leavevmode
\end{center}
\vspace{-0.3cm}
\caption[]{Diagram for the elastic production of a vector meson $V$ in
  $\gamma^* p$ collisions}
\label{fig1}
\end{figure}

In Fig.~\ref{fig1} the basic picture for the process $\gamma^* p \to V
p$ is shown: first the photon with virtuality $Q^2 = -q^2$ fluctuates into a
quark-antiquark pair. This $q\bar q$ fluctuation then interacts
elastically with the proton $p$, where the zig-zag line represents the
(for the moment unspecified) elastic interaction with the proton. The
$\gamma^* p$ centre-of-mass energy is denoted by $W$, 
\begin{equation}
W^2 = (q+p)^2\,,
\end{equation}
whereas
\begin{equation}
t = (p - p')^2
\end{equation}
is the four-momentum transfer squared. (In the following we will
mainly restrict ourselves to the case of small $t$.) The shaded blob at the
right stands for the formation of the vector meson $V$, which, to
leading order, has to form from the $q\bar q$ pair with invariant mass
squared $M_V^2$.
It is important to note that at high energy $W$ corresponding to
small values of $x$, 
\begin{equation}
x = \frac{Q^2 + M_V^2}{Q^2 + W^2}\,,
\label{eqdefx}
\end{equation}
the timescales involved in the problem are very
  different\footnote{This definition for $x$, which is often called
  $\xi$ or $x_{I\!\!P}$, is common in diffractive physics and should
  not be confused with the ordinary Bjorken-$x$, $x_{\rm Bj} =
  Q^2/(Q^2+W^2)$.}: the typical lifetime of the $\gamma^* \to q\bar q$
  fluctuation as well as the time for the formation of the vector
  meson $V$ is much longer than the duration of the interaction with the
  proton, i.e. $\tau_{\gamma^* \to q\bar q},\,\tau_{q\bar q \to V} \gg
  \tau_i$. Therefore the basic amplitude factorizes, as sketched already in
 Fig.~\ref{fig1}, into the $q\bar q$ fluctuation, the interaction
 amplitude $A_{q\bar q + p}$ and the wave function of the vector meson
 $V$,
\begin{equation}
A(\gamma^* p \to V p) \ = \ \psi^{\gamma}_{q\bar q} 
 \otimes A_{q\bar q + p} \otimes \psi_{q\bar q}^V\,,
\label{eqfac}
\end{equation}
and the process becomes calculable within various models.\footnote{For
  a more detailed discussion of the ordering of the timescales see
  e.g.\ \cite{LMRT}.} Formally it has been shown that for $Q^2$ larger
  than all other mass scales in the process there is factorization
  into a hard scattering subprocess, non-perturbative (and
  off-diagonal, as will be discussed later) parton distributions and
  the meson wave function.\cite{fac} This strict proof of
  factorization holds for longitudinally polarized photons, whereas
  meson production through transversely polarized photons is shown to
  be suppressed by a power of $Q$.

Let us now turn to the discussion of different models for the
$\gamma^* p$ interaction.

\section{Some non-perturbative models}
The following short section is far from being a review of this rich
field, but is intended to give a hint at some non-perturbative models,
which contrast the perturbative description of diffractive scattering,
which is the main subject of this article.\\[2mm]
$\bullet$ We will not cover approaches based on vector meson dominance (see
e.g.~\cite{Schildknechtetal}).\\[2mm] 
$\bullet$ For Regge-phenomenology-based models of (one or two) Pomeron
exchange we refer the reader to \cite{Peter}.\\[2mm]
$\bullet$ \emph{The model of the stochastic QCD vacuum}\\
Dosch, Gusset, Kulzinger and Pirner \cite{DGKP} have developped a
model of the interaction with the proton, which is similar to the
semi-classical model of Buchm\"uller discussed in \cite{Buchmuller} in
the context of inclusive diffractive DIS. This model, originally used
for hadron-hadron scattering, leads to linear confinement and predicts
a dependence of the high-energy scattering on the hadron size. It
gives a unified description of low energy and soft high-energy
scattering phenomena. Dosch et al.\ approximate the slowly
varying infrared modes of the gluon field of the proton by a
stochastic process. Via a path integral method they average over all
possible field configurations. For the splitting of the
photon into the $q\bar q$ pair and for the description of the vector
meson they use light cone wavefunctions. Within their model they are
able to calculate the $Q^2$ dependence of the cross section, as well
as the dependence on the momentum transfer $t$, ${\rm d}\sigma/{\rm d}
t$, and the ratio of the longitudinal to the transverse cross section,
$L/T$, where longitudinal and transverse refer to the polarization of
the photon. Their results are in fair agreement with experimental
data. There is no prediction for the $W$ dependence of the cross
section.\\[2mm]
$\bullet$ Rueter has extended the model of Dosch et al.\ to also describe the
$W$ dependence of the cross section.\cite{Rueter} He achieves this by
using a phenomenological model based on the exchange of one soft and
one hard Pomeron, each being a simple pole in the complex angular
momentum plane, similar to the Donnachie-Landshoff model
\cite{Peter}. For the very hard components of the photon fluctuations
he treats the interaction perturbatively and achieves a good
description of the experimentally observed transition from the soft to
the hard regime.

\section{The two gluon exchange model}
To leading order in QCD the zig-zag line in Fig.~\ref{fig1}, which
stands for the elastic scattering via the exchange of a colourless
object with the quantum numbers of the vacuum, can be described by two
gluons. If the scale governing the (transverse) size
of the photon fluctuation is large compared to the typical scale of
non-perturbative strong interactions, i.e.\ if
\begin{equation}
Q^2 \gg \Lambda_{\rm QCD}^2 \quad {\rm or} \quad 
M_V^2 \gg \Lambda_{\rm QCD}^2\,,
\end{equation}
then the coupling of the two gluons to the $q\bar q$ fluctuation can
be treated reliably within perturbative QCD (pQCD). Another kinematic regime,
where pQCD is applicable, is high-$t$ diffraction. There the hard
scale which is needed to ensure the validity of the perturbative
treatment is given by the large value of the momentum transfer
$t$, and one expects high-$t$ diffractin to be a good place to search
for the perturbative Pomeron.\cite{Forshaw} 

It has been shown some time ago that due to the factorization property
of the process the coupling of the two gluons to the proton
can, in the leading logarithmic approximation, be identified with the
ordinary (diagonal) gluon distribution in the
proton.\cite{Misha1,Brodskyetal,Bartelsetal} We will come back to this
point later when discussing the importance of off-diagonal gluon distributions.

\subsection{The basic formula}
The basic leading order formula for diffractive vector meson
production is given by \cite{Misha1,Brodskyetal} 
\begin{equation}
\frac{{\rm d} \sigma}{{\rm d}t}\left(\gamma^* p \to V p\right)
\Big|_{t=0} = \frac{\Gamma^V_{ee} M_V^3 \pi^3}{48 \alpha}
\frac{\alpha_s(\overline{Q}^2)^2}{\overline{Q}^8}
\left[x\,g(x,\overline{Q}^2)\right]^2
\left( 1+ \frac{Q^2}{M_V^2} \right)\,,
\label{eqleading}
\end{equation}
where $\alpha$ is the electromagnetic coupling and the gluon
distribution is sampled at the effective scale
\begin{equation}
\overline{Q}^2 = \left( Q^2 + M_V^2 \right)/4\,.
\label{eqdefqbarsq}
\end{equation}
In Eq.~(\ref{eqleading}) the non-relativistic approximation for the
vector meson wave function is used and the coupling of the vector
meson to the photon is encoded in the electronic width
$\Gamma^V_{ee}$. Note that Eq.~(\ref{eqleading}) is valid for
$t=0$. In the approach discussed in the following there is no
prediction for the $t$ dependence of the cross section, which is
assumed to be of the exponential form $\exp(-b|t|)$ with an
experimentally measured slope-parameter $b$, which may depend on the
vector meson $V$ and on $Q^2$. On the other hand, Eq.~(\ref{eqleading})
makes predictions for both the $Q^2$ and the $W$ dependence of the
cross section for longitudinally and transversely polarized photons for
all sorts of vector mesons, as long as either $Q^2$ or $M_V^2$ is
large enough to act as the hard scale. It is obvious that the $W$
dependence comes entirely from the gluon distribution
$x\,g(x,\overline{Q}^2)$, which enters quadratically in the cross section.

\subsection{Improvements beyond the leading order}
In the following we will discuss several improvements of the leading
order formula.\footnote{For more detailed discussions see
  e.g.~\cite{RRML,FKS} or the recent review \cite{MW}.}\\[2mm]

$\bullet$ Eq.~(\ref{eqleading}) contains only the leading imaginary
part of the positive-signature amplitude
\begin{equation}
A \propto i \left( x^{-\lambda} + (-x)^{-\lambda} \right)\,.
\end{equation}
The real part of the amplitude can be restored using dispersion
relations:
\begin{equation}
{\rm Re} A = \tan(\pi \lambda/2)\,{\rm Im} A\,,
\end{equation}
where $\lambda$ is given by the logarithmic derivative
\begin{equation}
\lambda = \frac{\partial \log A}{\partial \log(1/x)}\,.
\end{equation}
For the case of $\rho$ production, the contributions from the real
part are roughly 15\%. For $J/\Psi$ production in the HERA regime they
amount to approximateley 20\% and are even bigger for $\Upsilon$
production \cite{MRT3,FMS}, where larger values of $x$ are probed.\\[2mm]

\begin{figure}[htb]
\vspace{-0.5cm}
\begin{center}
\includegraphics[width=0.7\textwidth]{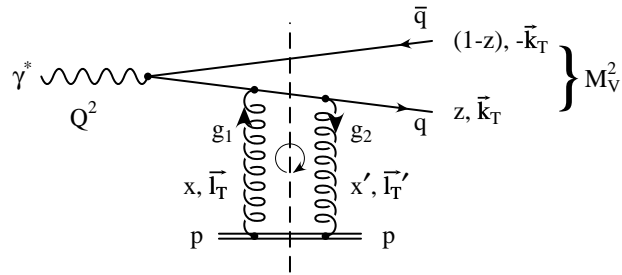}
\leavevmode
\end{center}
\vspace{-0.3cm}
\caption[]{One of four leading order diagrams for the two gluon
  exchange model for diffractive vector meson production}
\label{fig2}
\end{figure}

$\bullet$ In Fig.~\ref{fig2} one of four leading order
diagrams\footnote{There are three similar diagrams: one where both
  gluons couple to the antiquark and two where one gluon is attached to
  the quark, whereas the other couples to the antiquark.} for the
two gluon exchange model is shown with some kinematic variables which
will be used below. In the general case the two gluons $g_1$, $g_2$
have different $x$, $x^{\prime}$ and transverse momenta $\ell_T$,
$\ell_T^{\prime}$. The leading logarithmic approximation of the
$\ell_T^2$ loop integral (indicated by the circle in Fig.~\ref{fig2}) 
leads to the identification with the integrated gluon distribution $x\,
g(x, \overline{Q}^2)$ at the effective scale $\overline{Q}^2$ defined in
Eq.~(\ref{eqdefqbarsq}). Beyond leading logarithmic accuracy one has
to perform the $\ell_T^2$ integral over the unintegrated gluon
distribution $f(x,\ell_T^2)$. This can lead to numerical results which
are, depending on the kinematical regime, twice as big as the result
from Eq.~(\ref{eqleading}).\cite{RRML,LMRT} 

Although here we are considering elastic production at small momentum
transfer $t$, the timelike vector meson with mass $M_V$ has to be
produced from the spacelike (or real) photon with virtuality
$Q^2$. This means, that even if there is no transverse momentum transfer,
$\ell_T = \ell_T^{\prime}$, there has to be a difference $x -
x^{\prime} = \left(M_V^2 + Q^2\right)/\left(W^2 + Q^2\right)$ in the
longitudinal momentum of the two gluons $g_1$ and $g_2$. Therefore the
identification with the ordinary diagonal gluon distribution
$x\,g(x,\overline{Q}^2)$ is only a good approximation for very small
values of $x$ and $t$, and in the general case the process $\gamma^* p
\to V p$ depends on off-diagonal parton distributions.\cite{Alan}
Their importance for diffractive vector meson production will be
discussed in the following.

\section{Recent issues in pQCD calculations}
\subsection{Off-diagonal parton distributions}
Off-diagonal (also called ``skewed'' or non-forward) parton
distributions\footnote{These off-diagonal parton distributions are not
  parton densities in the ordinary probabilistic sense but matrix
  elements of parton-fields between different initial and final proton
  states.} are much studied recently.\footnote{See \cite{Alan} and
  references therein.} In the case of small $t$ scattering the
skewedness comes from the difference between $x$ and $x^{\prime}$ of
the two gluons $g_1$ and $g_2$, and the cross section can be shown to
be proportional to the square of a skewed gluon distribution,
\begin{equation}
\sigma \propto \left|x^{\prime} g\left(x, x^{\prime};
    \overline{Q}^2\right)\right|^2\,.
\label{eqsigmaskewed}
\end{equation}
Here
$x = \left(M_{q\bar q}^2 + Q^2\right)/\left(W^2 + Q^2\right)$, 
$x^{\prime} = \left(M_{q\bar q}^2-M_V^2\right)/\left(W^2 + Q^2\right)
\quad \ll \quad x$, and 
$M_{q\bar q}^2$ is the mass squared of the intermediate $q\bar
q$ pair. (Taking the leading imaginary part of the amplitude
corresponds to cutting the amplitude as indicated by the dashed line
in Fig.~\ref{fig2} and putting both $q$ and $\bar q$ on-shell, which
in turn fixes $x$. $x^{\prime}$ has to accomodate the difference
between $M_{q\bar q}$ and $M_V$ and it not fixed due to the
integration over all possible quark (and antiquark) momenta. At
leading logarithmic order $x^{\prime} \ll x$ and we can put
$x^{\prime} \simeq 0$.)

For arbitrary kinematics skewed parton distributions are not connected
with the diagonal ones and are unknown non-perturbative
objects. However, in the case of small $x$, they are determined
completely by the diagonal ones.\cite{SGMR,Alan} The ratio of skewed to
diagonal gluon distribution is given by
\begin{equation}
R_g \ = \ \frac{x^{\prime} g(x, x^{\prime})}{x\,g(x)} \ = \ 
\frac{2^{2\lambda+3}}{\sqrt{\pi}}\,\frac{\Gamma\left(\lambda +
    \frac{5}{2}\right)}{\Gamma\left(\lambda + 4\right)}\,.
\label{eqrg}
\end{equation}
Here $\Gamma$ is the usual Gamma-function and the effective power
$\lambda$ can be obtained from the logarithmic derivative of the amplitude
$A$ for the $\gamma^* p \to V p$ cross section,
\begin{equation}
\lambda = \frac{\partial \log A}{\partial \log\left(1/x\right)}\,.
\label{eqdeflambda}
\end{equation}
As will be shown below, the magnitude of the resulting correction
factor for the total cross section, $R_g^2$, can be sizeable,
especially for large $Q^2$ or $M_V^2$.

\subsection{The vector meson wave function}
Another important issue is the treatment of the vector meson wave
function. As sketched in Fig.~\ref{fig1} and Eq.~(\ref{eqfac}) it
enters the amplitude via a convolution with the scattered $q\bar q$
fluctuation. In Eq.~(\ref{eqleading}) the non-relativistic approximation
was adopted. This means, that quark and antiquark equally share the
longitudinal momentum of the photon, i.e.\ $\ z = 1-z = 1/2$, and that
there is no internal (transverse) momentum $k_T$ in the $q\bar q$
bound state. Therefore, in this naive approximation,
\begin{equation}
\psi^V_{q\bar q}(z, k_T) \ = \ 
\delta^{(2)}(k_T)\,\,\delta\left(z - 1/2\right)\,,
\label{eqnonrelwavefunction}
\end{equation}
and $M_V = 2m_q$. While this simplification may be suitable for heavy
mesons like the $\Upsilon$, it is clear that the non-relativistic
approximation has to break down for light quarks. Various groups have
worked on improving this approximation by including the Fermi motion
of the quarks in the meson by using a nontrivial wave
function.\cite{Brodskyetal,FKS,RRML,Nemchiketal} Different models for
the meson wave function were used which lead to quite different
correction factors: whereas in Gaussian models there is no strong
suppression \cite{RRML}, the large $k_T$ tail typical for wave
functions from non-relativistic potential models seems to lead to 
large corrections \cite{FKS}. On the other hand, considering that
within these potential models a big part of the ${\cal O}(v^2)$
corrections comes from a regime, where $k_T$ is bigger
than the quark mass itself, these large corrections may well be an
artefact of the non-relativistic approximation.

Another related problem is the question, which mass for the quarks
should be used in the perturbative formulae. Note that
Eq.~(\ref{eqleading}) is written in terms of the vector meson mass
$M_V$. However, as discussed in \cite{RRML}, the full expressions used
to include higher order (relativistic) corrections contain the quark
mass $m_q$ instead of $M_V$. As the ratio $M_V/(2m_q)$ enters with a
high power, this difference is not negligible and should be taken into
account in the calculation of the ${\cal O}(v^2)$ corrections applied
to Eq.~(\ref{eqleading}).

In addition, it is well known that there are other relativistic
corrections, which in principle have to be taken into account in a
consistent way. As pointed out by Hoodbhoy \cite{Hoodbhoy}, gauge
invariance is only preserved if higher Fock states ($q\bar q g$, $q\bar
q g g, \ldots$) are included in the wave function. In doing so he
arrives at the conclusion, that the relativistic corrections to the
quark propagators plus the corrections from the higher Fock states
amount to only a few percent for $J/\Psi$ production, in agreement
with \cite{RRML}.

After all large relativistic corrections can probably be excluded,
but, as different approaches lead to quite different results there
remains a considerable uncertainty and the issue a hot topic.

\subsection{An alternative approach based on parton-hadron duality}
In this section we will discuss an alternative approach, which avoids
the meson wave function and leads to results which are in surprisingly
good agreement with available data. It was proposed in \cite{MRT1} for
$\rho$ meson electroproduction, where the hard scale is provided by
$Q^2$, not by $M_{\rho}$.\footnote{Experimentally both the $t$
  dependence ${\rm d}\sigma/{\rm d}t \sim \exp(-bt)$ with $b \simeq
  5-6$ GeV$^{-2}$ for $Q^2 > 10$ GeV$^2$ and the $W$ behaviour of the
  cross section $\sigma \propto W^{0.8}$ indicate that $\rho$ meson
  electroproduction is not a soft, but mainly a hard process.}
 Due to the tiny $u$ and $d$ quark masses, in
the case of the $\rho$ non-relativistic approximations cannot be
justified, and the wave function is not very well known. Now the crucial
problem was, that all naive predictions for the ratio of the
longitudinal to the transverse cross section, which are based on the
perturbative formula (\ref{eqleading}), lead to 
\begin{equation}
\sigma_L/\sigma_T \sim Q^2/M_{\rho}^2\,.
\label{eqlovertnaiv}
\end{equation}
This is much too steep and incompatible with experimental data (see
below). The inclusion of effects from a light cone wave function for
the $\rho$ does not change the picture considerably.\footnote{One
  might argue that $\sigma_T$ receives large contributions from the
  small $k_T$ region, which is non-perturbative. But those
  contributions would cause the transverse cross section to fall off
  even faster with increasing $Q^2$ and therefore worsen the problem
  \cite{MRT1}.} These observations indicate that the main effects are
not coming from the $\rho$ wave function and lead to the proposal of
a different model in \cite{MRT1}: there the cross section for $\rho$
production is predicted via perturbative $u\bar u$ and $d\bar d$ quark
pair electroproduction together with the principle of parton-hadron
duality (PHD) \cite{PHD}. PHD means that the integral of the parton
($q\bar q$) production cross section over a mass interval $\Delta M$ is
approximately equal to the sum over all (corresponding) possible
hadron production cross sections in the same mass interval. In the
region $M_{q\bar q}^2 \approx M_{\rho}^2$ the production of more
complicated partonic configurations (like $q\bar q + g$, $q\bar q +
2g$, $q\bar q+q\bar q$, etc.) is heavily suppressed. On the hadronic
side the $\rho$ resonance (plus the small admixture of the $\omega$) with its
decay into two (three) pions completely saturates the cross
section. Therefore we can well approximate the $\rho$ production cross
section
$$
\gamma^* p \to \rho \, p \to \pi \pi\, p
$$
by
\begin{equation}
\sigma\left(\gamma^* p \to \rho \, p\right) \simeq 0.9\, \sum_{q = u, d}
\int_{M_a^2}^{M_b^2} \frac{{\rm d}\sigma\left(\gamma^* p \to (q\bar q)
    p \right)}{{\rm d}M^2}
\label{eqphdrho}
\end{equation}
where $M_a$ and $M_b$ have to be chosen to embrace the $\rho$
resonance appropriately, i.e.\ $M_b^2 - M_a^2 \sim 1$ GeV$^2$. The
factor 0.9 on the right side of Eq.~(\ref{eqphdrho}) corrects for the
contributions from $\omega$ production.

The perturbative formulae for the 
$q\bar q$ production cross section are derived from the amplitudes
depicted in Fig.~\ref{fig2} and can be written in terms of the
conventional spin rotation matrices $d_{\lambda \mu}^J (\theta)$ (see
\cite{MRT1} for details):
\begin{eqnarray}
\label{eqsigmalt}
\frac{{\rm d}^2 \sigma_L}{{\rm d}M^2 dt} \Big|_{t=0} & = & 
\frac{4 \pi^2 e_q^2 \alpha}{3} \: 
\frac{Q^2}{(Q^2 + M^2)^2} \: \frac{1}{8} \: \int_{-1}^1 {\rm d} \cos \theta \: 
\left | d_{10}^1 (\theta) \right |^2 \left | I_L \right |^2\,,\\
\frac{{\rm d}^2 \sigma_T}{{\rm d}M^2 dt} \Big|_{t=0} & = & 
\frac{4 \pi^2 e_q^2 \alpha}{3} \: 
\frac{M^2}{(Q^2 + M^2)^2} \: \frac{1}{4} \: \int_{-1}^1 {\rm d} \cos \theta \: 
\left ( \left | d_{11}^1 (\theta) \right |^2 + \left |d_{1-1}^1 
(\theta) \right |^2 \right ) \left | I_T \right |^2\nonumber
\end{eqnarray}
where $e_q$ is the electric charge of the quark $q$, $\alpha$ the
electromagnetic coupling and $\theta$ the polar angle of the quark $q$
in the $q\bar q$ rest frame with respect to the proton
direction ($k_T = M/2 \sin\theta$). $I_{L,T}$ are integrals over the gluon
$\ell_T^2$ and given by
\begin{eqnarray}
\label{eqdefilt}
I_L (K^2) & = & K^2 \: \int \: \frac{{\rm d}\ell_T^2}{\ell_T^4} \: \alpha_s 
(\ell_T^2) \: f (x, \ell_T^2) \: \left ( \frac{1}{K^2} \: - \: 
\frac{1}{K_\ell^2} \right )\,, \\
I_T (K^2) & = & \frac{K^2}{2} \: \int \: \frac{{\rm d}\ell_T^2}{\ell_T^4} \:
\alpha_s (\ell_T^2) \: f (x, \ell_T^2) \: \left ( \frac{1}{K^2} \: -
\: \frac{1}{2k_T^2} \: + \: \frac{K^2 - 2k_T^2 + \ell_T^2}{2k_T^2 \: 
K_\ell^2} \right )\,,\nonumber
\end{eqnarray}
with $f$ being the unintegrated gluon distribution and 
$$
K_\ell^2 \; \equiv \; \sqrt{(K^2 + \ell_T^2)^2 \: - \: 4k_T^2
  \ell_T^2}\,,\qquad K^2 \; \equiv \; k_T^2 (Q^2 + M^2)/M^2\,.
$$
In Eqs.~(\ref{eqsigmalt}) the different rotation matrices
appropriately reflect the different spin states of the $q\bar q$
produced from longitudinal and transverse photons, and the integrals
$I_{L, T}$ contain the scattering off the proton via the two gluon
exchange.\footnote{Here we assume $s$ channel helicity conservation
  (SCHC), i.e.\ the produced $\rho$ has the helicity of the virtual
  photon. However, there are small violations of SCHC
  (e.g. $\gamma^*_T \to \rho_L$) which can be
  successfully described in a framework similar to the one discussed
  here. For a discussion of recent measurements of the 15 spin
  density matrix elements of $\rho$ production compared to different
  theoretical predictions see \cite{AProskuryakov}.}
In order to pick up only those $u\bar u$, $d\bar d$ configurations
which correspond to the quantum numbers of the $\rho$, one has to project out
the $J^{PC} = 1^{--}$ states. This can be easily done on amplitude
level with the same rotation matrices $d_{\lambda \mu}^J (\theta)$,
see \cite{MRT1}. (Even higher spin states like the $\rho(3^-)$ can be
projected out using the corresponding $d$ function \cite{MRT2}.) It is
important to note that through the projection on amplitude level the
longitudinal and transverse cross sections $\sigma_{L, T}$ are less
infrared sensitive than Eqs.~(\ref{eqsigmalt}), and therefore
$\sigma_T$ becomes calculable without a large uncertainty from the
treatment of the (non-perturbative) infrared region.

For the complete numerical predictions one also has to include the
contributions from the real part of the amplitudes and the skewed
gluon distribution as discussed above. Both effects are taken into
account on amplitude level. Eqs.~(\ref{eqsigmalt}) give the cross
section differential in $t$ for $t=0$. To arrive at the $t$ integrated
total cross section one assumes the exponential form
$\exp(-b|t|)$. The slope $b$ can be taken from experiment or
theoretical models and depends in general on $M^2$, $W$ and $Q^2$. 
For more details we refer the reader to \cite{MRT4}.

To go beyond the leading order prediction in a completely consistent
way would require in addition the full set of next-to-leading order
gluonic corrections to the $(q\bar q)$-$2g$ vertex. These corrections
are not known yet\footnote{A first step towards the calculation of the
  full NLO corrections is provided by \cite{FM}.}, but can be
estimated by a ${\cal K}$ factor \cite{LMRT,MRT1}. Similar to the Drell-Yan
process, there are $\pi^2$ enhanced terms, which come from the $i\pi$
terms in the double logarithmic Sudakov form factor. Resummation of
those leading corrections results in the ${\cal K}$ factor
${\cal K} = \exp\left(\pi C_F \alpha_s\right)$
which leads to a considerable enhancement of the cross section.

\begin{figure}[htb]
\vspace{-0.8cm}
\begin{center}
\includegraphics[width=0.7\textwidth]{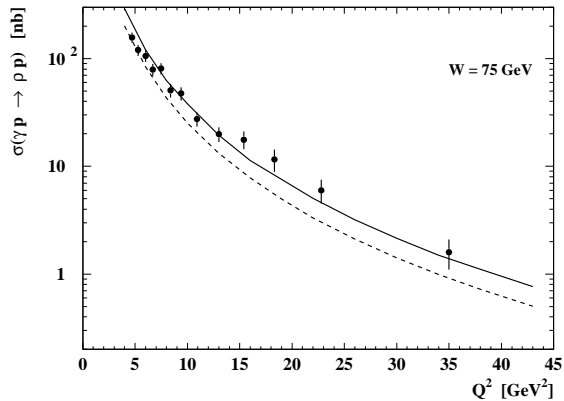}
\leavevmode
\end{center}
\vspace{-0.8cm}
\caption[]{$\sigma\left(\gamma^* p \to \rho \, p\right)$ predicted
  within the PHD model as described in the text compared to recent H1
  data \cite{H1}. {\it Continuous line}: using the skewed gluon
  distribution, {\it dashed line}: without skewing}
\label{fig3}
\end{figure}
In Fig.~\ref{fig3} the complete numerical prediction for $\gamma^* p
\to \rho \, p$ using the PHD model\footnote{For the numerical analysis the
  MRST99 gluon \cite{MRST99} was used and the scale of $\alpha_s$ in
  the ${\cal K}$ factor was chosen as $2K^2$. For more details see
  \cite{MRT4}.}
is shown as a function of $Q^2$ together with recent H1 data.\cite{H1}
The continuous line includes all the effects discussed above, whereas the
dashed line does not include the skewed gluon. The importance of the
off-diagonal gluon for the $Q^2$ behaviour of the cross section is
obvious and the effect seems to be required to describe the data. Of
course the model prediction is not free from
uncertainties like the choice of the mass interval $M_b^2 - M_a^2$ in
Eqs.~(\ref{eqphdrho}) or the scale of $\alpha_s$ in the ${\cal K}$
factor. These (and other) uncertainties are discussed in detail in
\cite{MRT4}, but they affect mainly the normalization of the cross
section and do not spoil the good agreement with the experimental data.
\begin{figure}[htb]
\vspace{-0.8cm}
\begin{center}
\includegraphics[width=0.6\textwidth]{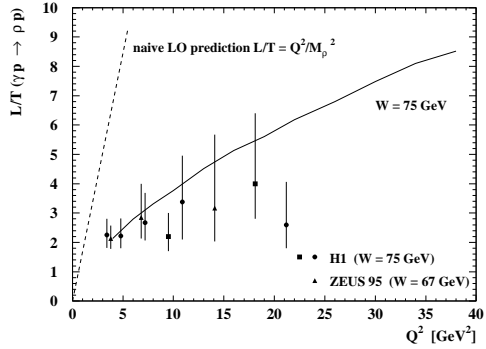}
\leavevmode
\end{center}
\vspace{-0.8cm}
\caption[]{$L/T$ predicted within the PHD model ({\it continuous
    line}) as described in the text compared to experimental data
  \cite{H1,H1old,ZEUS} and
  the naive prediction ({\it dashed line}) from Eq.~(\ref{eqlovertnaiv})}
\label{fig4}
\end{figure}
In Fig.~\ref{fig4} the prediction of the PHD model for the ratio $L/T$
is shown as a continuous line. It agrees fairly well with the data
points, which show a very modest rise with $Q^2$ in contrast to the
naive prediction Eq.~(\ref{eqlovertnaiv}) (dashed line). Thus, in the
PHD picture, it is {\em not} the $\rho$ wave function, but the
dynamics of the $q\bar q$ pair creation from longitudinal and
transverse photons together with the off-diagonal two gluon
interaction and the projection onto the $1^-$ state, that determines
the $Q^2$ dependence and the ratio $L/T$.

\begin{figure}[htb]
\vspace{-0.8cm}
\begin{center}
\includegraphics[width=0.6\textwidth]{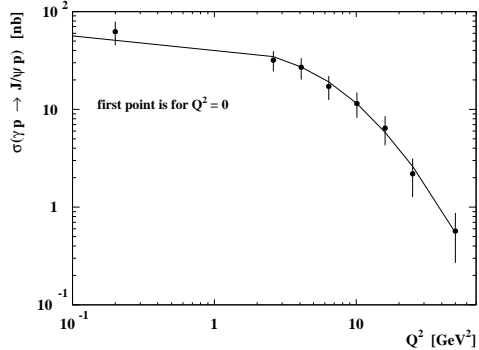}
\leavevmode
\end{center}
\vspace{-0.8cm}
\caption[]{Cross section for diffractive $J/\Psi$ production as predicted
  in the PHD model \cite{MRT4} compared to recent H1 data \cite{H1two}}
\label{fig5}
\end{figure}
It is important to note that the PHD model also works in the case of
massive quarks and heavy mesons. Starting from formulae for diffractive heavy
quark production \cite{LMRT} and modifying the projection formalism
appropriately, elastic $\Upsilon$ photoproduction was recently
predicted using PHD in agreement with first measurements.\cite{MRT3} The
same formalism can also be applied to diffractive $J/\Psi$
production.\cite{MRT4} Again, as shown in Fig.~\ref{fig5}, there is a
surprisingly good agreement between the predicted cross section as a
function of $Q^2$ and the experimental data \cite{H1two}.

\section{Summary}
Elastic vector meson production is a rich field, both from 
experimental and theoretical point of view. Different theoretical
models describe the data, and more precise data in an increased kinematical
range will be needed to clarify the situation. We have briefly
discussed some non-perturbative models, but mainly concentrated on
perturbative approaches. We have shown that
with recent improvements pQCD-based approaches work very well and are in
agreement with data. The fairly large impact of skewed parton
distributions on the predictions within these models is supported by
the data. There is good hope that in future we will be able to
discriminate between the different models and to understand elastic
vector meson production in more detail. By combining different
observables from different processes elastic vector meson production
with its high sensitivity to the gluon at small $x$ will finally help
to constrain the gluon much better. For this much effort will be
needed also from the theoretical side in order to increase the
precision of the calculations.

\subsection*{Acknowledgements}
I would like to thank G. Grindhammer, B. Kniehl and G. Kramer for the
good organization of this stimulating and enjoyable workshop. I also
thank Genya Levin, Alan Martin and Misha Ryskin for pleasant
collaborations.

\end{document}